\documentstyle[pre,aps,preprint,epsfig]{revtex} 
\begin{document}                                     
\title{Social force model for pedestrian dynamics}
\author{Dirk Helbing and P\'{e}ter Moln\'{a}r}
\address{II. Institute of Theoretical Physics, University of
Stuttgart, 70550 Stuttgart, Germany}                           
\maketitle                                                    
\begin{abstract}                                           
It is suggested that the motion of pedestrians can be described as if they
would be subject to `social forces'. 
These `forces' are not directly exerted by the
pedestrians' personal environment, but they are a measure for the internal
motivations of the individuals to perform certain actions (movements).
The corresponding force concept is discussed in more detail
and can be also applied to the description of other behaviors.
\par
In the presented model of pedestrian behavior several force terms
are essential: First, a term describing the acceleration towards the desired
velocity of motion. Second, terms reflecting that a pedestrian 
keeps a certain distance
to other pedestrians and borders. Third, a term modeling attractive effects.
The resulting equations of motion are 
nonlinearly coupled {\sc Langevin} equations. 
Computer simulations of crowds 
of interacting pedestrians show that the social force model 
is capable of describing the self-organization of several
observed collective effects of pedestrian behavior very realistically.
\end{abstract}
\section{Introduction}

During the last two decades models of pedestrian behavior have found notable
interest for several reasons. First, there are some striking analogies with
gases \cite{Hel1,Hel2,Hel3} and fluids \cite{Hel1,Hel2,Hen1,Hen2}. 
Second, all model quantities like
places $\vec{r}_\alpha$ and velocities $\vec{v}_\alpha$ of pedestrians
$\alpha$ are measurable and, therefore, comparable with empirical 
data.
Third, there already exists a considerable amount of data material like flow
measurements or (video) films \cite{Oed,Nav,HCM}. 
Fourth, pedestrian models can provide
valuable tools for designing and planning pedestrian areas, subway or railroad
stations, big buildings, shopping malls, etc. 
\cite{Gipps,Borgers,Lovas}.
\par
In the 1970s {\sc Henderson} \cite{Hen1,Hen2} has compared
measurements of pedestrian flows with {\sc Navier-Stokes} equations with
considerable success. His phenomenological 
fluid-dynamic approach has been improved 
and mathematically founded by
{\sc Helbing} in the 1990s on the basis of a pedestrian specific
gas\-kinetic ({\sc Boltz\-mann}-like) model \cite{Hel1,Hel2}. 
This model is closely related to some gaskinetic \cite{Prig,Pav} and
fluid-dynamic \cite{Kern1,Kern2,PRE} {\em traffic} models. 
Recently, much attention has been attracted by {\em `microscopic'}
approaches for vehicular traffic \cite{Biham,Nagel,Nagat,Cuesta}.
In the following, we will introduce a {\em social force model} for pedestrian
motion. First ideas of this `microscopic' modeling concept were 
formulated in \cite{Hel2,Hel5}. A short overview of 
`microscopic', `mesoscopic' and `macroscopic' descriptions of pedestrian
behavior and their interrelations is given in Ref. \cite{Hel6}.

\section{The social force concept}

Many people have the feeling that human behavior is `chaotic' or at least very
irregular and not predictable. This is probably true for behaviors that are
found in complex situations. However, at least for 
relatively simple situations {\em stochastic}
behavioral models may be developed if one restricts to the
description of behavioral {\em probabilities} that can be found 
in a huge population (resp. group) of 
individuals (see \cite{Hel3,WeHa,Weidl}). This idea has been followed
by the {\em gaskinetic} pedestrian model \cite{Hel1,Hel2,Hel3}.
\par
Another approach for modeling behavioral changes has been suggested by
{\sc Lewin} \cite{Lewin}. According to his idea behavioral changes are
guided by socalled {\em social fields} resp.\ {\em social forces}.
In the following we will examine if this idea could be applied to the
description of pedestrian behavior. 
\par
Figure 1 illustrates a scheme of the
processes that lead to behavioral changes. 
According to this, a sensory
{\em stimulus} causes a behavioral             
{\em reaction} that depends on the personal
aims and is chosen from a set of behavioral alternatives with the
objective of utility maximization. 
\par
Table 1 suggests a classification of stimuli
into simple or standard situations that are well predictable, and complex
or new situations that may be modelled with probabilistic models.
However, since a pedestrian is used to the
situations he/she is normally confronted with, 
his/her reaction is usually rather
automatic, and determined by his/her experience of which reaction will be
the best. It is therefore possible to put the rules of pedestrian behavior
into an equation of motion. According to this equation 
the systematic temporal changes
$d\vec{w}_\alpha/dt$ 
of the {\em prefered velocity}
$\vec{w}_\alpha(t)$ 
of a pedestrian 
$\alpha$ 
are described by a vectorial quantity 
$\vec{F}_\alpha(t)$ 
that can be
interpreted as {\em social force}. Clearly, this force must represent the
effect of the environment (e.g. other pedestrians or borders) on
the behavior of the described pedestrian. However, 
the social force is not excerted by the environment on a 
pedestrian's body. It is rather a quantity that describes
the concrete {\em motivation to act}. In the case of pedestrian behavior
this motivation evokes the physical production of an acceleration or
deceleration force as reaction to the perceived
information that he/she obtains about his/her environment (see figure 1). 
In summary, one can say that a pedestrian acts {\em as if} he/she would
be subject to external forces. This idea has been mathematically founded in 
Ref. \cite{Hel4}.

\section{Formulation of the social force model}

In the following the main effects that determine the motion of a 
pedestrian $\alpha$ will be introduced:
\begin{itemize}
\item[1.] He/She wants to reach a certain 
destination $\vec{r}_\alpha^{\,0}$ as comfortable
as possible. Therefore, he/she normally takes a way 
without detours, i.e., the
shortest possible way. This way will usually have the shape of a polygon with
edges $\vec{r}_\alpha^{\,1},\dots,\vec{r}_\alpha^{\,n} 
:= \vec{r}_\alpha^{\,0}$. If
$\vec{r}_\alpha^{\,k}$ is the next edge of this polygon to reach, his/her 
{\em desired direction} $\vec{e}_\alpha(t)$ of motion will be
\begin{equation}
 \vec{e}_\alpha(t) := \frac{\vec{r}_\alpha^{\,k} - \vec{r}_\alpha(t)}{\|
 \vec{r}_\alpha^{\,k} - \vec{r}_\alpha(t) \|} \, ,
\end{equation}
where $\vec{r}_\alpha(t)$ denotes the {\em actual position} 
of pedestrian $\alpha$ at time $t$.
Exactly speaking, the goals of a pedestrian are usually rather {\em gates} or
{\em areas} than points $\vec{r}_\alpha^{\,k}$. In this case, he/she will
at every time $t$ steer for the {\em nearest} point $\vec{r}_\alpha^{\,k}(t)$ 
of the corresponding gate/area.
\par
If a pedestrian's motion is not disturbed, he/she will walk into the desired
direction $\vec{e}_\alpha(t)$ 
with a certain {\em desired speed} $v_\alpha^0$. A
deviation of the {\em actual velocity} $\vec{v}_\alpha(t)$ from the
{\em desired velocity} $\vec{v}_\alpha^{\,0}(t) 
:= v_\alpha^0 \vec{e}_\alpha(t)$
due to necessary deceleration processes or avoidance processes leads to a
tendency to approach $\vec{v}_\alpha^{\,0}(t)$ 
again within a certain {\em relaxation
time} $\tau_\alpha$. This can be described by an {\em acceleration term}
of the form
\begin{equation}
 \vec{F}_\alpha^{\,0}(\vec{v}_\alpha,v_\alpha^0\vec{e}_\alpha)
 := \frac{1}{\tau_\alpha} (v_\alpha^0 \vec{e}_\alpha - \vec{v}_\alpha ) .
\end{equation}
\item[2.] The motion of a pedestrian $\alpha$ 
is influenced by other pedestrians. 
Especially, he/she keeps a certain distance from other pedestrians
that depends on
the pedestrian density and the desired speed $v_\alpha^0$.
Here, the {\em privat sphere} 
of each pedestrian, which can be interpreted as 
{\em territorial effect} \cite{Ash}, plays an 
essential role. A pedestrian normally
feels increasingly incomfortable the closer he/she gets to a strange person, 
who may react in an aggressive way. 
This results in {\em repulsive effects} of other pedestrians $\beta$
that can be represented by vectorial quantities
\begin{equation}
 \vec{f}_{\alpha\beta}(\vec{r}_{\alpha\beta}) := -
\nabla_{\vec{r}_{\alpha\beta}} V_{\alpha\beta} [b(\vec{r}_{\alpha\beta})] \, .
\end{equation}
We will assume that the repulsive potential $V_{\alpha\beta}(b)$ is a
monotonic decreasing function of $b$
with equipotential lines having the form
of an ellipse that is directed into the direction of motion. 
The reason for this is that 
a pedestrian requires space for 
the next step which is taken into account
by other pedestrians. $b$ denotes the semi-minor axis
of the ellipse and is given by
\begin{equation}
 2b := \sqrt{ (\|\vec{r}_{\alpha\beta}\| + \|\vec{r}_{\alpha\beta} - v_\beta
 \, \Delta t \, \vec{e}_\beta \| )^2 - ( v_\beta \, \Delta t )^2  } \, ,
\label{for4}
\end{equation}
where $\vec{r}_{\alpha\beta} := \vec{r}_\alpha - \vec{r}_\beta$. $s_\beta
:= v_\beta \, \Delta t$ is of the order of  
the step width of pedestrian $\beta$.
Despite the simplicity of this approach, it describes avoidance maneuvers
of pedestrians quite realistically.
\par
A pedestrian also keeps a certain distance 
from {\em borders} of buildings, walls, streets, obstacles, etc. \cite{HCM}. 
He/She feels the more
incomfortable the closer to a border he/she walks since he/she has to pay
more attention to avoid the danger of getting hurt, e.g.
by accidentally touching a wall. 
Therefore, a border $B$ evokes a {\em repulsive effect} that can
be described by 
\begin{equation}
 \vec{F}_{\alpha B} (\vec{r}_{\alpha B}) := - \nabla_{\vec{r}_{\alpha B}}
 U_{\alpha B} (\|\vec{r}_{\alpha B} \|) 
\end{equation}
with a repulsive and monotonic decreasing potential $U_{\alpha
B}(\|\vec{r}_{\alpha B}\|)$. Here, the vector $\vec{r}_{\alpha B} := 
\vec{r}_\alpha - \vec{r}_B^{\,\alpha}$ has been introduced, 
where $\vec{r}_B^{\,\alpha}$ denotes the location of that
piece of border $B$ that is nearest to pedestrian $\alpha$. 
\item[3.] Pedestrians are sometimes attracted by other persons (friends,
street artists, etc.) or objects (e.g. window displays). These {\em attractive
effects} $\vec{f}_{\alpha i}$ at places $\vec{r}_i$ 
can be modelled by attractive, monotonic increasing potentials
$W_{\alpha i}(\|\vec{r}_{\alpha i}\|,t)$ in a similar way like the repulsive
effects:
\begin{equation}
 \vec{f}_{\alpha i}(\| \vec{r}_{\alpha i}\|,t) := - \nabla_{\vec{r}_{\alpha i}}
 W_{\alpha i}(\|\vec{r}_{\alpha i} \|,t) 
\end{equation}
($\vec{r}_{\alpha i} := \vec{r}_\alpha - \vec{r}_i$).
The main difference is that the {\em attractiveness} $\|\vec{f}_{\alpha i}\|$
is normally decreasing with time $t$
since the interest is declining. The attractive effects are, e.g.,
responsible for the formation of pedestrian groups (that are comparable to
molecules).
\end{itemize}
However, the formulas above for attractive and repulsive effects only hold
for situations that are perceived in the desired direction $\vec{e}_\alpha(t)$
of motion. Situations 
located behind a pedestrian will have a weaker influence $c$ with $0 < c < 1$.
In order to take this effect of perception (i.e. of the {\em effective 
angle $2\varphi$ of sight})
into account we have to introduce the direction dependent weights
\begin{equation}
 w(\vec{e},\vec{f}) := \left\{
\begin{array}{ll}
1 & \mbox{if } \vec{e} \cdot \vec{f} \ge \| \vec{f} \| \cos \varphi  \\
c & \mbox{otherwise.}
\end{array}\right.
\end{equation}
In summary, the 
repulsive and attractive effects on a pedestrian's behavior
are given by
\begin{eqnarray}
 \vec{F}_{\alpha\beta}(\vec{e}_\alpha,\vec{r}_\alpha - \vec{r}_\beta)
 &:=& w(\vec{e}_\alpha,-\vec{f}_{\alpha\beta})\vec{f}_{\alpha\beta}
 (\vec{r}_\alpha - \vec{r}_\beta) \, , \nonumber \\
 \vec{F}_{\alpha i}(\vec{e}_\alpha,\vec{r}_\alpha - \vec{r}_i,t)
 &:=& w(\vec{e}_\alpha,\vec{f}_{\alpha i})\vec{f}_{\alpha i}
 (\vec{r}_\alpha - \vec{r}_i,t) \, . 
\end{eqnarray}                               
We can now set up the equation for a pedestrian's total 
motivation $\vec{F}_\alpha(t)$. Since all
the previously mentioned effects influence a pedestrian's decision at
the same moment, we will assume that their total effect is given by the
sum of all effects, like this is the case for forces.
This results in
\begin{eqnarray}
 \vec{F}_\alpha(t)
 &:=& \vec{F}_\alpha^{\,0}(\vec{v}_\alpha,
 v_\alpha^0\vec{e}_\alpha) 
 + \sum_\beta \vec{F}_{\alpha \beta} (\vec{e}_\alpha,
 \vec{r}_\alpha - \vec{r}_\beta) \qquad \qquad \nonumber \\
 &+& \sum_B \vec{F}_{\alpha B}(\vec{e}_\alpha,\vec{r}_\alpha 
     - \vec{r}_B^{\,\alpha})
 + \sum_i \vec{F}_{\alpha i}(\vec{e}_\alpha,\vec{r}_\alpha - \vec{r}_i,t) \, .
\label{dynforce}
\end{eqnarray}
The {\em social force model} is now defined by
\begin{equation}
 \frac{d\vec{w}_\alpha}{dt} := \vec{F}_\alpha(t) + \mbox{\em fluctuations.} 
\label{sozforce}
\end{equation}
Here, we have added a {\em fluctuation term} that takes into account
random variations of the behavior. These
fluctuations stem, on the one hand, from ambiguous situations in which two or
more behavioral alternatives are equivalent (e.g. if 
the utility of passing an obstacle 
on the right or left hand side is the same). On the other
hand, fluctuations arise from accidental or deliberate deviations from the
usual rules of motion.
\par  
In order to complete the model of pedestrian dynamics a relation between the
actual velocity $\vec{v}_\alpha(t)$ and the 
prefered velocity $\vec{w}_\alpha(t)$ must be introduced.
Since the actual speed is limited by a pedestrian's {\em maximal acceptable
speed} $v_\alpha^{\rm max}$, we will assume that 
the {\em realized} motion is given by
\begin{equation}
 \frac{d\vec{r}_\alpha}{dt} = \vec{v}_\alpha(t) 
 := \vec{w}_\alpha(t) \, g \left( 
 \frac{v_\alpha^{\rm max}}{\| \vec{w}_\alpha \|} \right)
\label{sozdyn}
\end{equation}
with
\begin{equation}
 g\left( \frac{v_\alpha^{\rm max}}{\| \vec{w}_\alpha \|} \right)
 := \left\{
\begin{array}{ll}
 1 & \mbox{if } \| \vec{w}_\alpha \| \le v_\alpha^{\rm max} \\
 v_\alpha^{\rm max} / \| \vec{w}_\alpha \| & \mbox{otherwise.}
\end{array}\right.
\label{ge}
\end{equation}
\par
Note that the pedestrian model (\ref{sozforce}), (\ref{sozdyn})
has the form of nonlinearly coupled {\sc Langevin} equations. 
A simplified version of it can be extended to an 
{\em active walker model} \cite{Lam1,Lam2,Schw} for the self-organization of
systems of ways \cite{SFB}. This is related to the mathematical
description of trunk trail formation by ants \cite{Frank}.

\section{Computer simulations}

The model of pedestrian dynamics developed in Section III
has been simulated on a computer
for a large number of interacting pedestrians confronted with different
situations. Despite the fact that the the proposed model is very simple it
describes a lot of observed phenomena very realistically. In the following,
two examples will be presented 
showing the {\em self-organization} of collective
phenomena of pedestrian behavior.
\par
The simulations assumed that the desired speeds $v^0$ are {\sc Gaussian}
distributed \cite{Hen1,Nav} with mean $\langle v^0 \rangle 
= 1.34\mbox{ms}^{-1}$ and standard deviation $\sqrt{\theta} 
= 0.26\mbox{ms}^{-1}$ \cite{Weidmann}. 
Speeds were limited to $v_\alpha^{\rm max} = 1.3
v_\alpha^0$. Pedestrians enter the walkway at the ends at random
positions. Those intending to walk from the left
to the right hand side are represented by full circles, whereas pedestrians 
intending to move into the opposite direction are represented by empty circles.
The diameter of a circle is a measure for the actual speed $\|\vec{v}_\alpha
\|$ of a pedestrian $\alpha$.
For simplicity, no attractive effects
$\vec{f}_{\alpha i}$ or fluctuations were taken into account. 
The repulsive potentials 
were assumed to decrease exponentially, i.e.
\begin{equation}
 V_{\alpha\beta}(b) = V_{\alpha\beta}^0 \mbox{e}^{-b/\sigma} \, , \qquad
 U_{\alpha B}(\| \vec{r}_{\alpha B} \|) 
  = U_{\alpha B}^0 \mbox{e}^{-\| \vec{r}_{\alpha B} \| / R}
\end{equation}
with $V_{\alpha\beta}^0 = 2.1 \mbox{m}^2\mbox{s}^{-2}$, $\sigma =
0.3\mbox{m}$ and $U_{\alpha B}^0 = 10 \mbox{m}^2\mbox{s}^{-2}$, $R =
0.2\mbox{m}$. 
Potentials with
a hard core would be more realistic, of course, but they will not yield other
results since the simulated pedestrians always keep enough distance. 
\par
Walkways were chosen 10m wide. For $\Delta t$ in formula (\ref{for4}) we took 
$\Delta t = 2 \mbox{s}$, and for the relaxation times we used
$\tau_\alpha = 0.5 \mbox{s}$. Smaller values of $\tau_\alpha$ let the
pedestrians walk more aggressive. Finally, the effective angle of sight (which
also takes into account head movements) was set to $2\varphi = 200^\circ$.
Situations outside the angle of sight were assumed to have an influence of $c=
0.5$. The model parameters introduced above
were chosen in a way that is compatible with empirical data.
\par
Figure 2 shows the empirically confirmed \cite{Oed} development of 
dynamically varying {\em lanes}
consisting of pedestrians who intend 
to walk into the same direction. 
Periodic boundary conditions in transversal direction would 
stabilize these lanes since they were not any more destroyed at the
ends of the walkway by randomly entering pedestrians \cite{Hel2,Hel5}. Figure
3 shows the number of forming lanes in dependence of the width of the
walkway for a pedestrian density of $0.3\mbox{m}^{-2}$.  
\par
The segregation effect of lane
formation is {\em not} a result of the initial pedestrian 
configuration but a consequence of the pedestrians' interactions.
Nevertheless, it normally leads to a more effective pedestrian flow
since time-consuming avoidance maneuvers occur less frequently.
\par
Figures 4 and 5 depict different moments of two pedestrian groups 
that try to pass a narrow door into opposite directions. 
The corresponding simulation shows the following:
Once a pedestrian has passed the door, other pedestrians intending to move
into the same direction are able to follow him/her easily
(see Figure 4).
However, 
the stream of passing pedestrians
is stopped by the pressure of the opposing group after some time.
Subsequently, the door is captured by pedestrians who pass the door into 
the opposite direction (see Figure 5). This                
change of the passing direction
may occur several times and is well-known from observations.

\section{Summary and outlook}

It has been shown that pedestrian motion
can be described by a simple social force model for individual
pedestrian behavior. Computer simulations of pedestrian groups demonstrated 
1.\ the development of lanes consisting of pedestrians who walk into the
same direction, 2.\ oscillatory changes of the walking direction at
narrow passages. These spatio-temporal patterns arise due to the {\em nonlinear
interactions} of pedestrians. They are {\em not} the effect of strategical
considerations of the individual pedestrians since they were assumed to
behave in a rather `automatical' way. 
\par
Presently, the social force model is extended by a
model for the route choice behavior of pedestrians. 
As soon as the computer program is completed 
it will provide a comfortable tool for town- and traffic-planning.
\par
The investigation of pedestrian behavior
is an ideal starting point for the development of other or
more general quantitative behavioral
models, since the 
variables of pedestrian motion are easily measurable so that
corresponding models are comparable with empirical data.
A further step could be the application of the social force concept 
to the description of opinion formation, 
group dynamics, or other social phenomena \cite{Hel4}. For this purpose,
an abstract behavioral space has to be introduced.

\section*{Acknowledgements}

The authors want to thank W. Weidlich, 
M. R. Schroeder, and W. Scholl for valuable and inspiring discussions.

\clearpage
\begin{table}
\begin{center}
\begin{tabular}{lll}
{\it Stimulus} & Simple/Standard & Complex/New \\
               & Situations      & Situations \\[1ex]
{\it Reaction} & Automatic Re-   & Result of Evaluation, \\                 
               & action, `Reflex' & Decision Process \\[1ex]
{\it Characterization} & Well Predictable & Probabilistic \\[1ex]
{\it Modeling} & Social Force  & Decision Theore- \\
{\it Concept}  & Model, etc.     & tical Model, etc. \\[1ex]
{\it Example}  & Pedestrian      & Destination Choice \\
               & Motion          & by Pedestrians 
\end{tabular}
\end{center}
\caption{Classification of behaviors according to their complexity.}
\end{table}
\clearpage
{\small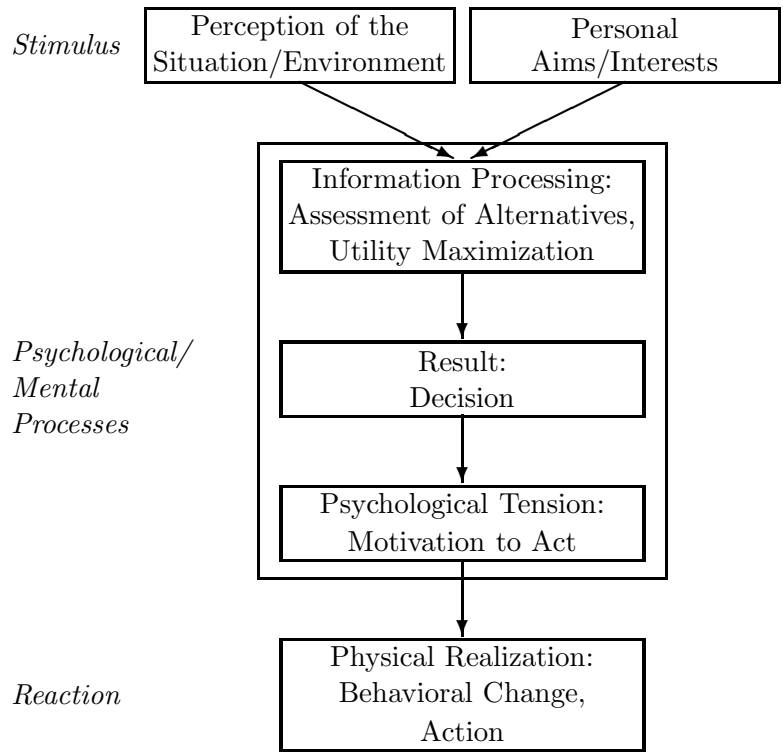
\begin{figure}
\unitlength0.6cm
\begin{center}
\begin{picture}(17,16.4)(-4,0)
\thicklines
\put(2,0){\framebox(8,2.4){}}
\put(6,1.6){\makebox(0,0.8){Physical Realization:}} 
\put(6,0.8){\makebox(0,0.8){Behavioral Change,}}
\put(6,0){\makebox(0,0.8){Action}}
\put(1.5,3.8){\framebox(9,9.6){}}                    
\put(6,4.2){\vector(0,-1){1.7}}
\put(2,4.2){\framebox(8,1.6){}}                                         
\put(6,4.2){\makebox(0,0.8){Motivation to Act}}
\put(6,5){\makebox(0,0.8){Psychological Tension:}}
\put(6,7.4){\vector(0,-1){1.5}}
\put(2,7.4){\framebox(8,1.6){}}                              
\put(6,7.4){\makebox(0,0.8){Decision}}
\put(6,8.2){\makebox(0,0.8){Result:}}
\put(6,10.6){\vector(0,-1){1.5}}
\put(2,10.6){\framebox(8,2.4){}}
\put(6,10.6){\makebox(0,0.8){Utility Maximization}}
\put(6,11.4){\makebox(0,0.8){Assessment of Alternatives,}}
\put(6,12.2){\makebox(0,0.8){Information Processing:}}       
\put(2.4,14.8){\vector(2,-1){3.4}}
\put(-1,14.8){\framebox(6.8,1.6){}}
\put(2.4,14.8){\makebox(0,0.8){Situation/Environment}}                
\put(2.4,15.6){\makebox(0,0.8){Perception of the}}
\put(9.6,14.8){\vector(-2,-1){3.4}}
\put(6.2,14.8){\framebox(6.8,1.6){}}
\put(9.6,14.8){\makebox(0,0.8){Aims/Interests}}
\put(9.6,15.6){\makebox(0,0.8){Personal}}
\put(-4,15.4){{\it Stimulus}}
\put(-4,1.0){{\it Reaction}}
\put(-4,7.0){{\it Processes}}                              
\put(-4,7.8){{\it Mental}}
\put(-4,8.6){{\it Psychological/}}
\end{picture}
\end{center}
\caption[]{Schematic representation of processes leading to behavioral 
changes.\label{fig1}}
\end{figure}}
\clearpage
\unitlength1cm
\begin{figure}[htbp]
\begin{center}
\begin{picture}(16,10.6)(0,-0.8)
\put(0,9.8){\epsfig{height=16\unitlength, width=9.8\unitlength, angle=-90, 
      bbllx=50pt, bblly=50pt, bburx=554pt, bbury=770pt, 
      file=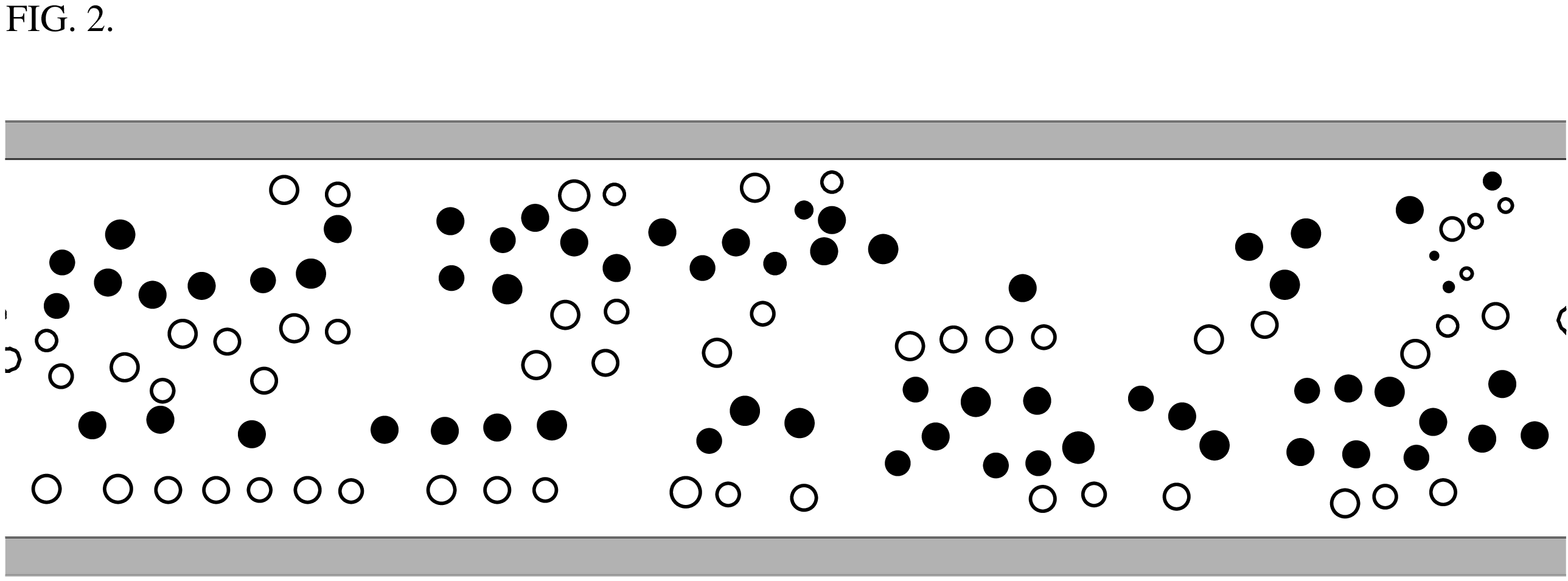}}
\end{picture}
\end{center}
\caption[]{Above a critical pedestrian density one can observe the
formation of lanes consisting of pedestrians with a uniform walking
direction. Here, the computational result shows $N = 4$ lanes on a walkway
that is 10m wide and 50m long. Empty circles
represent pedestrians with a desired direction of motion which is
opposite to that of pedestrians symbolized by full circles.\label{fig2}}
\end{figure}
\clearpage
\begin{figure}[htbp]
\begin{center}
\begin{picture}(16,10.6)(0,-0.8)
\put(0,9.8){\epsfig{height=16\unitlength, width=9.8\unitlength, angle=-90, 
      bbllx=50pt, bblly=50pt, bburx=554pt, bbury=770pt, 
      file=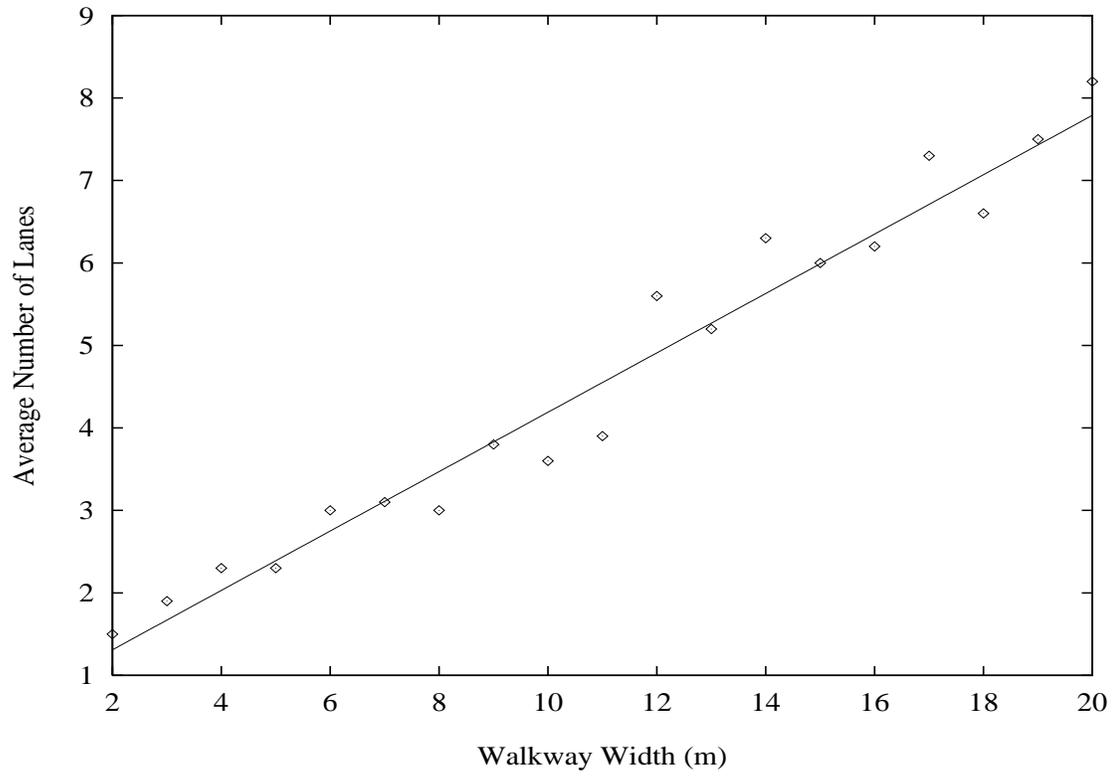}}
\end{picture}
\end{center}
\caption[]{The average number $N$ of lanes emerging on a walkway 
scales linearly with its width $W$ 
($N(W) = 0.36\mbox{m}^{-1} W + 0.59$).\label{fig3}}
\end{figure}
\clearpage
\begin{figure}[htbp]
\begin{center}
\begin{picture}(16,10.6)(0,-0.8)
\put(0,9.8){\epsfig{height=16\unitlength, width=9.8\unitlength, angle=-90, 
      bbllx=50pt, bblly=50pt, bburx=554pt, bbury=770pt, 
      file=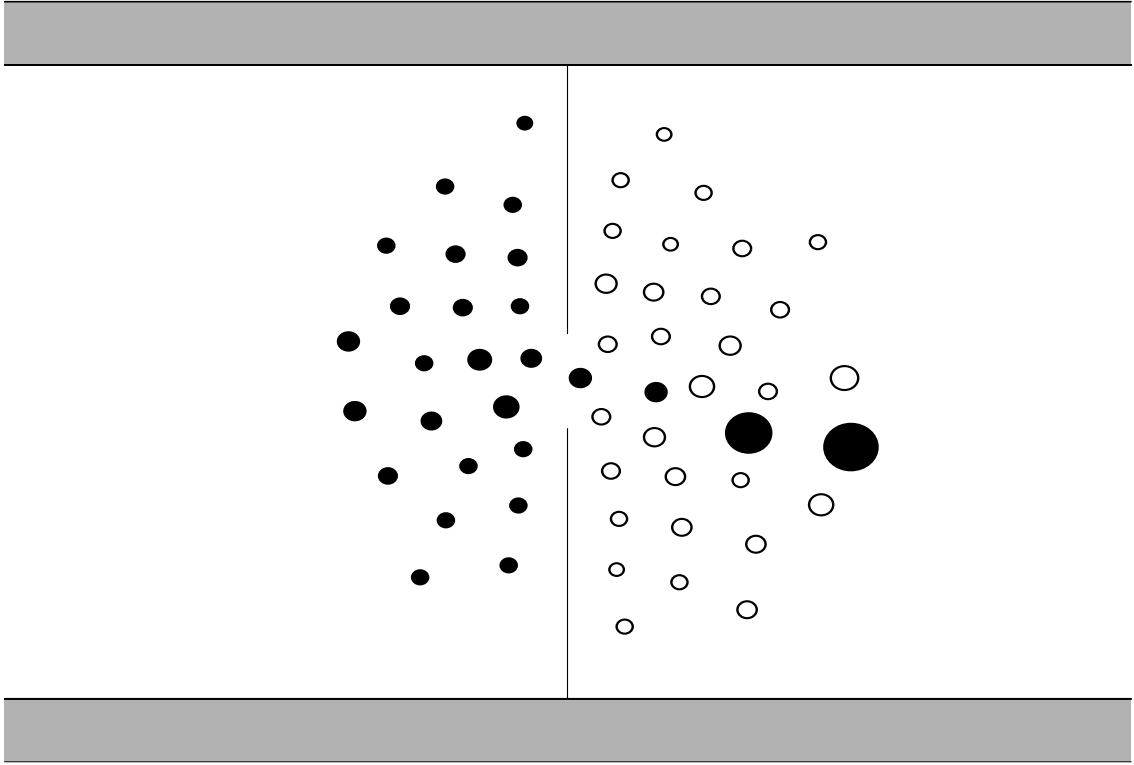}}
\end{picture}
\end{center}
\caption[]{If one pedestrian has been 
able to pass a narrow door, other pedestrians
with the same desired walking direction can follow easily whereas
pedestrians with an opposite desired direction of motion have to wait. 
The diameters of the
circles are a measure for the actual velocity of motion.\label{fig4}}
\end{figure}
\clearpage
\begin{figure}[htbp]
\begin{center}
\begin{picture}(16,10.6)(0,-0.8)
\put(0,9.8){\epsfig{height=16\unitlength, width=9.8\unitlength, angle=-90, 
      bbllx=50pt, bblly=50pt, bburx=554pt, bbury=770pt, 
      file=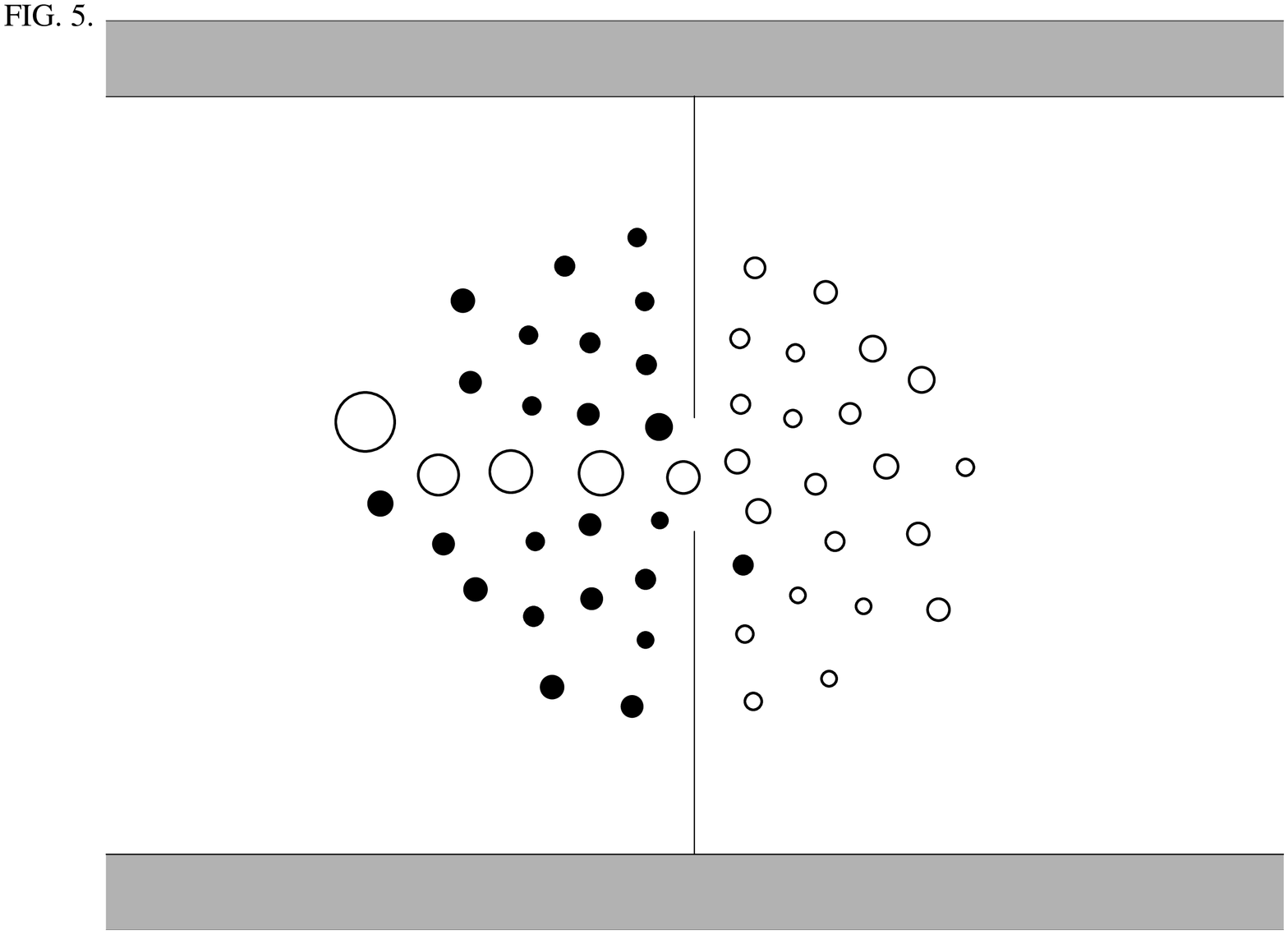}}
\end{picture}
\end{center}
\caption[]{After some time the pedestrian stream is stopped, and the door is
captured by individuals who pass the door into the opposite
direction.\label{fig5}}
\end{figure}

\begin{references}
\bibitem{Hel1} D. Helbing, {Complex Systems} {\bf 6}, 391 (1992).
\bibitem{Hel2} D. Helbing, {\it Physikalische Modellierung des dynami\-schen
Verhaltens von Fu\ss{}g\"angern} (Diplom thesis, Georg-August University,
G\"ottingen, Germany, 1990).
\bibitem{Hel3} D. Helbing, {\em Stochastische Methoden, nichtlineare Dynamik
und quantitative Modelle sozialer Prozesse} (Shaker, Aachen, Germany, 1993).
\bibitem{Hen1} L. F. Henderson, {Nature} {\bf 229}, 381 (1971).
\bibitem{Hen2} L. F. Henderson, {Transp. Res.} {\bf 8}, 509 (1974).
\bibitem{Oed} D. Oeding, {\it Verkehrsbelastung und Dimensionierung von
Gehwegen und anderen Anlagen des Fu\ss{}g\"angerverkehrs}
(Stra\ss{}enbau und Stra\ss{}enverkehrstechnik, Heft 22,
Bonn, Germany, 1963).
\bibitem{Nav} F. P. D. Navin and R. J. Wheeler, {Traffic Engineering}
{\bf 39}, 30 (1969).
\bibitem{HCM} {\em Highway Capacity Manual}, Chap. 13 (Transportation
Research Board, Special Report 209, Washington, D.C., 1985).
\bibitem{Gipps} G. P. Gipps and B. Marksj\"o, {Math. Comput. Simul.}
{\bf 27}, 95 (1985).
\bibitem{Borgers} A. Borgers and H. Timmermans, {Geographical Analysis}
{\bf 18}, 115 (1986).
\bibitem{Lovas} G. G. L\o{}v\aa{}s, in Proceedings of the 1993 European
Simulation Multiconference, Lyon, France, June 7--9, 1993.
\bibitem{Prig} I. Prigogine and R. Herman, {\it Kinetic Theory of Vehicular
Traffic} (American Elsevier, New York, 1971). 
\bibitem{Pav} S. L. Paveri-Fontana, Transp. Res. {\bf 9}, 225
(1975).
\bibitem{Kern1} B. S. Kerner and P. Konh\"auser, Phys. Rev. E {\bf 48},
2335 (1993).
\bibitem{Kern2} B. S. Kerner and P. Konh\"auser, Phys. Rev. E {\bf 50}, 54
(1994).
\bibitem{PRE} D. Helbing, An improved fluid-dynamic model for vehicular
traffic, Phys. Rev. E. (submitted).
\bibitem{Biham} O. Biham, A. A. Middleton, and D. Levine, Phys. Rev. A
{\bf 46}, 6124 (1992).
\bibitem{Nagel} K. Nagel and M. Schreckenberg, J. Phys. I France {\bf 2},
2221 (1992).
\bibitem{Nagat} T. Nagatani, Phys. Rev. E {\bf 48}, 3290 (1993).
\bibitem{Cuesta} J. A. Cuesta, F. C. Mart\'{\i}nez, J. M. Molera, and
A. S\'{a}nchez, Phys. Rev. E. {\bf 48}, 4175 (1993).
\bibitem{Hel5} D. Helbing, {Behavioral Science} {\bf 36}, 298 (1991).
\bibitem{Hel6} D. Helbing, in {\it Natural Structures. Principles,
Strategies, and Models in Architecture and Nature}, Part II
(Sonderforschungsbereich 230, Stuttgart, Germany, 1992).
\bibitem{WeHa} W. Weidlich and G. Haag, {\it Concepts and Models of a
Quantitative Sociology} (Springer, Berlin, 1983).
\bibitem{Weidl} W. Weidlich, {Physics Reports} {\bf 204}, 1 (1991).
\bibitem{Lewin} K. Lewin, {\it Field Theory in Social Science} 
(Harper \& Brothers, New York, 1951).
\bibitem{Hel4} D. Helbing, {Physica A} {\bf 196}, 546 (1993).
\bibitem{Ash} A. E. Scheflen and N. Ashcraft, {\it Human Territories:
How We Behave in Space-Time} (Prentice-Hall, Englewood Cliffs, 1976).
\bibitem{Lam1} R. D. Freimuth and L. Lam, in {\it Modeling Complex Phenomena},
edited by L. Lam and V. Naroditsky (Springer, New York, 1992).
\bibitem{Lam2} D. R. Kayser, L. K. Aberle, R. D. Pochy, and L. Lam,
{Physica A} {\bf 191}, 17 (1992).
\bibitem{Schw} F. Schweitzer and L. Schimansky-Geier, {Physica A} {\bf 206},
359 (1994). 
\bibitem{SFB} D. Helbing, P. Moln\'{a}r, and F. Schweitzer,
in {\it Evolution of Natural Structures} 
(Sonderforschungsbereich 230, Stuttgart, Germany, 1994).
\bibitem{Frank} F. Schweitzer, K. Lao, and F. Family, Active Walker simulate
trunk trail formation by ants, Adaptive Behavior (submitted).
\bibitem{Weidmann} U. Weidmann
{\it Transporttechnik der Fu\ss{}g\"anger,} pp. 87--88 (Schriftenreihe
des Instituts f\"ur Verkehrsplanung, Transporttechnik, Stra\ss{}en- und
Eisenbahnbau Nr. 90, ETH Z\"urich, 1993).
\end{references}
\end{document}